\documentclass{INTERSPEECH2023}
\usepackage{hyperref}

\interspeechcameraready

\usepackage{todonotes}
\usepackage{multirow}
\usepackage{caption, subcaption}

\title{Expressive Machine Dubbing \\Through Phrase-level Cross-lingual Prosody Transfer}
\name{Jakub Swiatkowski$^*$\thanks{$^*$ Equal contribution}, Duo Wang$^*$, Mikolaj Babianski$^*$, Giuseppe Coccia,  Patrick Lumban Tobing, Ravichander Vipperla, Viacheslav Klimkov, Vincent Pollet}
\address{Amazon Science \email{\{jswiat|duowangd|babiansk|gcoccia|patrilum|ravivip|vklimkov|vinpolle\}@amazon.com}}

\begin{document}

\maketitle
 
\begin{abstract}
Speech generation for machine dubbing adds complexity to conventional Text-To-Speech solutions as the generated output is required to match the expressiveness, emotion and speaking rate of the source content. 
Capturing and transferring details and variations in prosody is a challenge. We introduce phrase-level cross-lingual prosody transfer for expressive multi-lingual machine dubbing. The proposed phrase-level prosody transfer delivers a significant 6.2\% MUSHRA score increase over a baseline with utterance-level global prosody transfer, thereby closing the gap between the baseline and expressive human dubbing by 23.2\%, while preserving intelligibility of the synthesised speech.

\end{abstract}

\section{Introduction}

Prosody transfer is the ability to transfer speaking style variations and vocal performances disentangled from the spoken content and speaker identity~\cite{googlee2e,wang2018style,luong2017adapting,hsuhierarchical,8683623}. Many of recent proposed methods utilize global-level prosody transfer. A single embedding per utterance is used to capture prosody and to condition the models to generate speech with the target prosody. These global embeddings are either explicitly learned from ground truth labels such as emotions~\cite{luong2017adapting,guo2022emodiff,rattcliffe22_interspeech}  or implicitly learned from a reference audio signal using a reference prosody encoder~\cite{googlee2e,wang2018style,8683623}, or a combination of both~\cite{hsuhierarchical}.

In this work, we focus on prosody transfer for cross-lingual machine dubbing. We aim to generate speech for a translated text in a target language with expressiveness and emotion of speech from multimedia content in source language. 
Exploration of cross-lingual prosody transfer is scarce.
\cite{rattcliffe22_interspeech} studied cross-lingual style transfer based on categorical labels, but this limits transfer of a wide range of expressions and emotions present in multimedia content. Until recently, existing work on machine dubbing has generated speech from translated text only, without transferring prosody~\cite{Federico2020, matouvsek2012improving, effendi2022duration}. To our best knowledge, VIPT (Variational Inference for Prosody Transfer)~\cite{anonymous2023cross} is the only known work tackling cross-lingual prosody transfer for machine dubbing. VIPT introduces learning of cross-lingual prosody transfer without parallel datasets using a VITS-based system~\cite{kim2021conditional} with a global reference encoder to capture vocal performance. One limitation in using a global reference embedding is that only utterance-level prosody variations are captured, while detailed local prosody variations %
cannot be properly encoded. This potentially impacts the transfer of prosody for generating long-form utterances. Transferring local prosody variations such as syntactic phrasing, topic emphasis and marked tonicity is important for expressive machine dubbing~\cite{torresquintero2021adept}.

In this work, we tackle the aforementioned drawback by exploring more fine-grained cross-lingual prosody transfer. Intra-lingual fine-grained prosody control has been explored~\cite{renfastspeech,lee2019robust,10022793,9053520} by training predictors of prosody components at phoneme or at word level. However, prosody transfer across different utterances at  word level suffers due to word mismatch. This applies as well to cross-lingual prosody transfer for machine dubbing, as it is unlikely to have one-to-one alignment between words in the original and translated text. Several recent works~\cite{9414966,9747158} have proposed machine translation techniques for machine dubbing allowing to generate monotonic alignments between translated texts at the level of a prosodic phrase, where a prosodic phrase is defined as a continuous segment of speech separated by silence regions. Therefore, in this work we explore phrase-level cross-lingual prosody transfer for machine dubbing.

Prosody delivery varies across languages, however the prosody of speech expressing the same emotions is correlated in related languages, as discussed in Section 4.6 in~\cite{brannon2022dubbing}. In our work, we explore these cross-lingual correlations for the purpose of prosody transfer. Our study is limited to European languages comprising English, German, French, Italian and Spanish, and focused on English-Spanish prosody transfer as a common dubbing language pair. We anticipate that more distant language pairs such as English-Japanese exhibit less correlated prosody features.

Our solution follows VIPT~\cite{anonymous2023cross} in combining a VAE (Variational Auto-Encoder) prosody encoder with VITS and trains on multimedia data without mining of parallel utterances with matching text across different locales. We propose to capture and to transfer phrase-level variations of prosody in a cross-lingual setting. To achieve that, we have devised a new phrase-level reference encoder that learns to condition the phrases of the input text with prosody embeddings extracted from corresponding parts of the reference speech waveform. We have also introduced a novel regularization applied on prosody embeddings based on phrase length, to reduce content leakage from short phrases. We discuss the details in Section~\ref{sec:method}. %

We evaluate our proposed method with both MUSHRA~\cite{recommendation2015bs} subjective perceptual test and objective metrics including Word Error Rate (WER) and conditional Fréchet DeepSpeech Distance (cFDSD)~\cite{binkowski2019high}. We compare our method against VIPT~\cite{anonymous2023cross}, a strong baseline for cross-lingual performance transfer. Both subjective and objective metrics suggest that our method improves expressiveness without compromising on intelligibility. We also show the importance of phrase-level conditioning in training, by comparing against a VIPT variant trained with global-level conditioning, but transferring prosody at phrase-level during inference. We demonstrate a significant 6.2\% MUSHRA score increase over VIPT, which closes the gap between machine dubbing and expressive human dubbing by 23.2\%. To summarize, our contributions are:
\begin{itemize}
    \item We present a new method capable of cross-lingual phrase-level prosody transfer for expressive multi-lingual machine dubbing. Robust and more fine-grained transfer compared to global-level prosody transfer improves the quality.
    \item We propose a length-based regularization method for fine-grained prosody representations.
\end{itemize}

\begin{figure*}[!h]
\begin{subfigure}[b]{.6\linewidth}
  \centering
  \includegraphics[width=10cm]{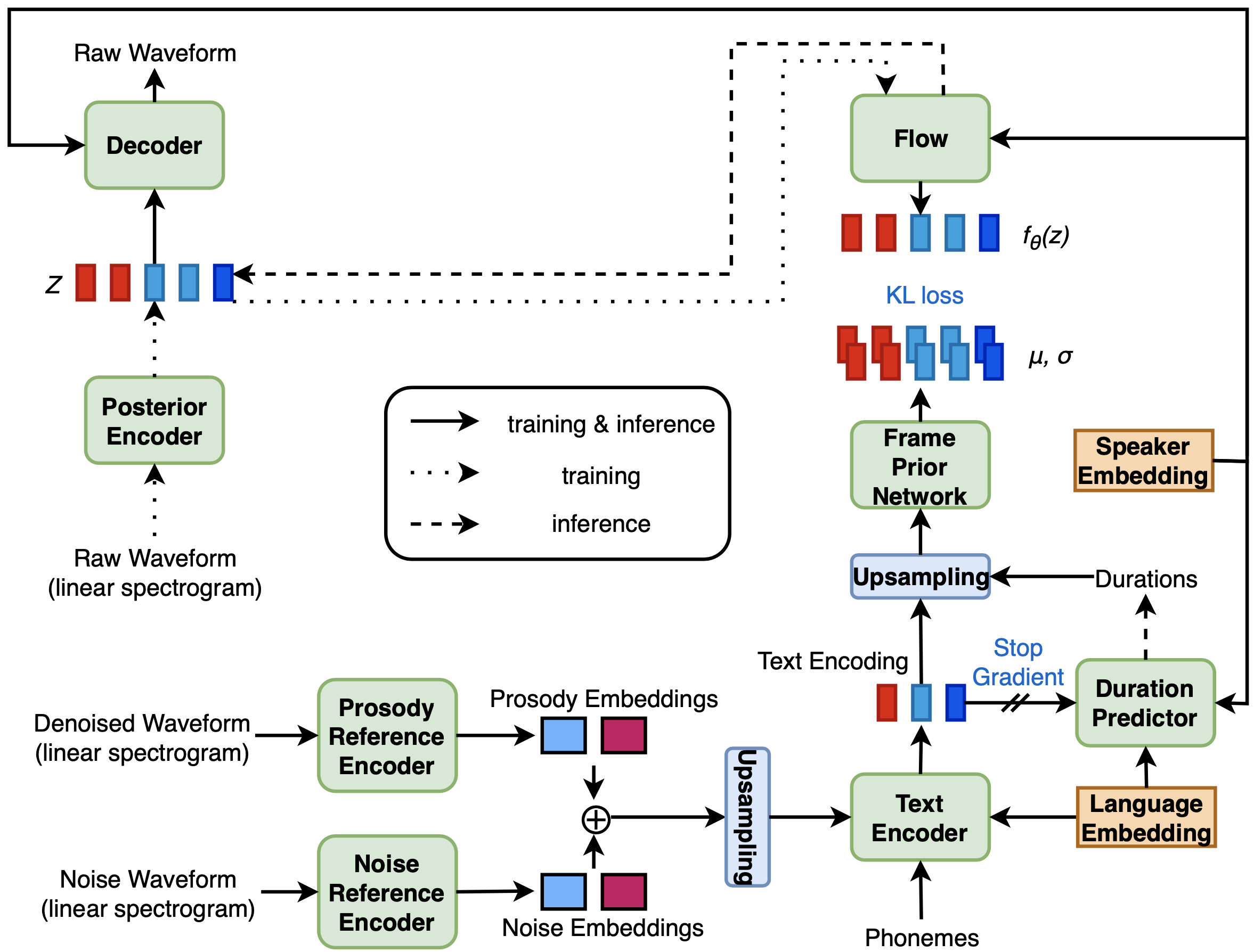}
  \caption{Proposed system architecture}
  \label{fig:overview}
\end{subfigure}
\hfill
\begin{subfigure}[b]{0.4\linewidth}
  \centering
  \includegraphics[width=5.7cm]{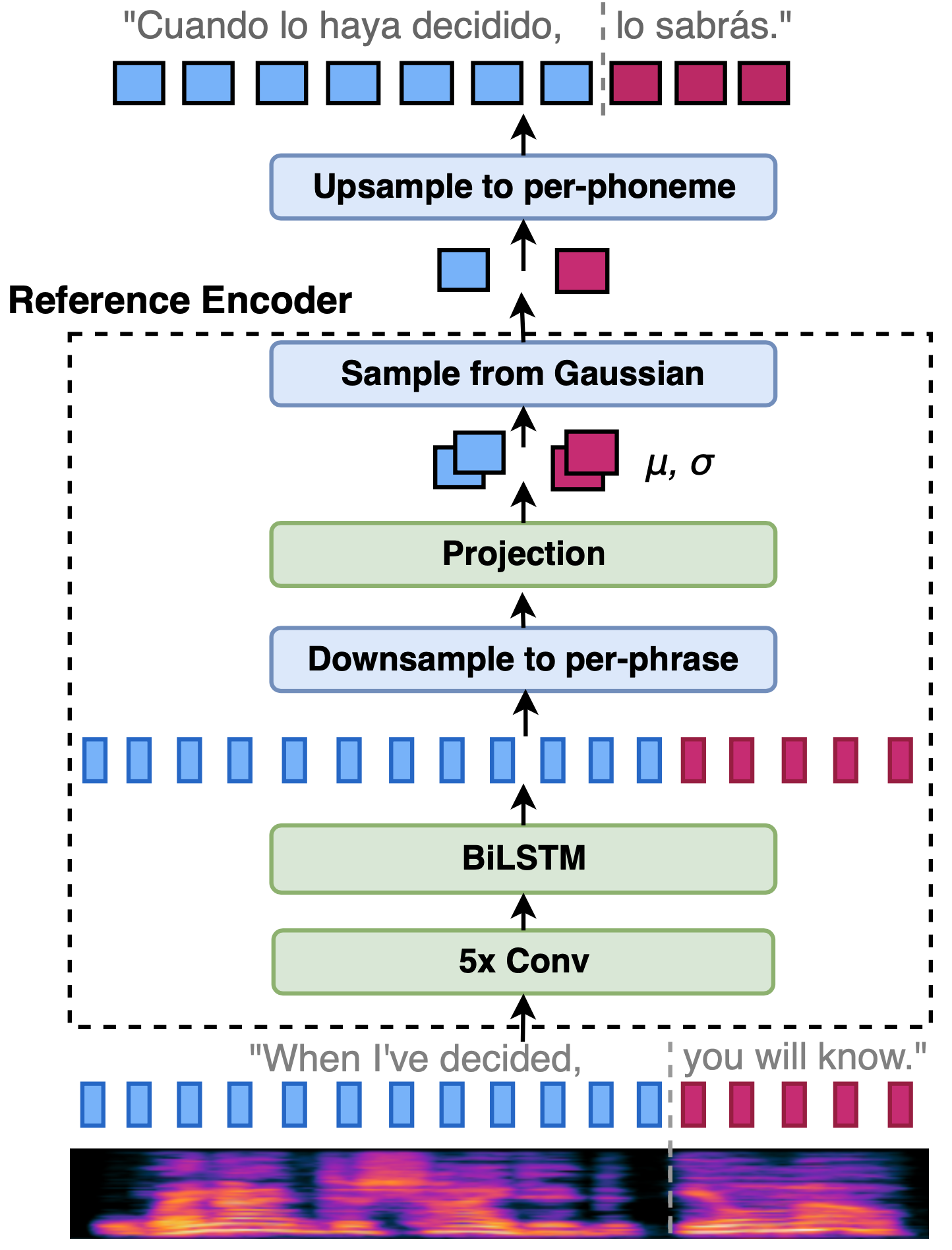}
  \caption{Phrase-level reference encoder}
  \label{fig:ref_enc_arch}
\end{subfigure}
\caption{Overview of the proposed system architecture and phrase-level reference encoder.}
\vspace{-3mm}
\end{figure*}

\section{Method}\label{sec:method}

This work extends the VIPT~\cite{anonymous2023cross} architecture by enabling modelling and cross-lingual transfer of prosody at phrase level. Figure~\ref{fig:overview} provides an overview of the proposed method. This section summarizes the baseline VIPT architecture and describes the extensions that enable the modelling and cross-lingual prosody transfer at the phrase level.

\subsection{VIPT architecture}
The VIPT architecture is based on conditional variational autoencoder with adversarial learning for end-to-end text-to-speech (VITS)~\cite{kim2021conditional}. VIPT makes a number of architectural changes to VITS. Most notably, VIPT combines VITS with an audio reference encoder that enables cross-lingual prosody transfer at global level. Consequently, VIPT is able to learn the cross-lingual prosody transfer from non-parallel data. This is possible as VIPT learns prosody representations that are agnostic to speakers and languages. Such representations can be transplanted from a reference audio in source language spoken by a source speaker to generate speech in the target language with the voice characteristics of a target speaker.

Furthermore, VIPT proposes a noise modelling variant of the reference encoder that allows clean speech synthesis even when training with noisy data and transferring prosody from noisy reference audio, which is common in multimedia data recorded in the field. 
The noise modelling variant consists of two separate reference encoders for denoised and noise streams obtained from the reference audio using a denoiser component \cite{Isik2020}. The reference encoders produce global-level denoised prosody and noise embeddings. These embeddings are then concatenated and broadcasted to phoneme-level in a prior encoder module. The VIPT architecture is analogous to ours shown in Figure~\ref{fig:overview}, but with a single global-level embedding per reference encoder instead of phrase-level embeddings. 

Finally, VIPT adapts a number of additional changes from the literature to the base VITS model. First, VIPT replaces VITS's monotonic alignment search algorithm with an explicit duration predictor and extends the prior encoder module with a frame prior network as in \cite{zhang2022visinger}. Second, VIPT adds speaker embeddings and language embeddings as in \cite{cho2022sane}.  Lastly, VIPT replaces HiFiGAN decoder \cite{kong2020hifi} with a BigVGAN-based decoder \cite{lee2022bigvgan}. We keep these components as in VIPT.

\subsection{Phrase-level cross-lingual prosody transfer}
We extend VIPT by enabling cross-lingual transfer of prosody at phrase level. The global-level reference encoder in VIPT may not sufficiently transfer the prosody variations present in expressive multimedia speech, especially when a dialogue line consists of several short phrases. Capturing prosody variations consequently requires  more fine-grained representations. Word-level prosody representations are challenging to transfer in a cross-lingual setting due to lack of monotonic word correspondence between translated texts.
Instead, we propose prosodic phrases as a level of granularity for cross-lingual prosody transfer.
We show experimentally that prosodic phrases are able to capture local variations in prosody which can be robustly transferred between speech in different languages.
At the same time, prosodic phrases can be automatically aligned across translated texts using recently developed prosodic alignment techniques for machine dubbing~\cite{9414966,9747158}.

\subsubsection{Phrase-level reference encoder}\label{method:ref_encoder}

We adopt the definition of prosodic phrases from~\cite{9414966,9747158} as continuous speech segments separated by silences.
The silences are extracted by force aligning reference audio and text using an external aligner, such as the Gaussian Mixture Model (GMM) based Kaldi Speech Recognition Toolkit~\cite{Povey_ASRU2011} used in our experiments.
We treat the silence regions as part of preceding speech phrases. Each speech phrase is encoded into a single prosody embedding using a reference encoder described below. See Figure~\ref{fig:ref_enc_arch} for an illustration of this approach. %

We propose a phrase-level reference encoder architecture that extracts frame-level embeddings from a linear spectrogram and downsamples the frame-level embeddings to the phrase-level.
We base our reference encoder architecture on that from VIPT, but with changes to keep one-to-one frame-embedding correspondence before downsampling to the phrase-level. Namely, our architecture consists of five convolutional layers with channel size of 512, a kernel size of three and stride of one, followed by one bi-directional LSTM layer with channel size of 512. The frame-level outputs of the bi-directional LSTM are then downsampled by selecting the middle embedding per phrase. We experimented with other forms of downsampling (e.g. mean of frames per phrase), but did not observe significant differences. The phrase-level embeddings are then further processed by a fully connected layer that outputs a parameterization of a 32-dimensional diagonal Gaussian distribution, which is regularized using Kullback-Leibler Divergence (KLD) with a standard Gaussian $\mathcal{N}(0, I)$. The final phrase embeddings are sampled from this Gaussian.

\subsubsection{Length-based regularization}\label{mehtod:length_based_regularization}
We propose a length-based regularization of phrase-level prosody embeddings to reduce content leakage when transplanting prosody embeddings across languages.
As we have observed content leakage for short phrases, we added a length-based regularization as following: 
\begin{equation}\label{eq:phrase_kld}
\mathcal{L}_{KLD_{\beta}} = \frac{1}{K} \sum_{k\in[1,K]} e^{-\beta L_{k}} KLD(h_k, \mathcal{N}(0,I))
\end{equation}
where $K$ is the total number of phrases in one utterance, $L_{k}$ is length of phrase $k$ defined as the number of phonemes, $\beta$ is a constant hyper-parameter controlling how much  $L_{k}$ affects the scaling factor $e^{-\beta L_{k}}$,  $h_k$ is the prosody embedding distribution of the $k^{th}$ phrase. %
This formulation applies stronger regularization to the embeddings of short phrases, thus preventing them from carrying content information that should only be obtained from the text prior. We show experimentally that the proposed regularization improves the intelligibility of synthesised speech for short phrases in Table~\ref{table:objective_metrics}.

\subsubsection{Noise modelling at phrase-level}
We apply the noise modelling approach from VIPT in the context of the phrase-level reference encoder. Specifically, the reference audio is passed to a denoising component~\cite{Isik2020} to extract denoised and noise streams. The two streams are then used as inputs to two separate phrase-level reference encoders with the architecture described in subsection~\ref{method:ref_encoder}. The reference encoders output phrase-level denoised prosody and noise embeddings that are concatenated, upsampled to the phoneme level and used as conditioning in the text encoder.

During inference, we extract a clean noise embedding from a static clean audio similarly to VIPT. However, we make sure to use a clean audio, which contains exactly one phrase, so that it is feasible to upsample this single clean noise embedding to match the number of the per-phrase prosody embeddings extracted from the denoised reference audio.

\subsubsection{Alignment of phrase-level audio reference embeddings to target text phonemes}
We aim to transfer prosody from phrases of reference speech to corresponding phrases in the translated target text to be synthesised.
More precisely, we concatenate phrase embeddings extracted from the reference audio with encoded phonemes corresponding to a given phrase in the target text. To achieve this, we need a mapping between reference audio phrases and target text phonemes. See the illustration in Figure~\ref{fig:ref_enc_arch}.

During training, the reference audio and the text to be synthesised correspond to each other. This allows us to force align the audio and the text phonemes to compute frame-phoneme correspondences. 
During inference, when performing cross-lingual prosody transfer, the reference audio contains speech in a language  different from the translated text to be synthesised. Therefore, in such case we cannot align the audio and the text as in training. Instead, we need to insert phrase breaks into the translated text to obtain a monotonic one-to-one alignment between the phrases in the audio and the text.
Such alignments can be automatically generated using recently developed prosodic alignment techniques for machine dubbing~\cite{9414966,9747158}. However, the focus of this work is on evaluating the quality of prosody transfer, and thus we assume the prosodic alignment is given.

\section{Experiments}

\subsection{Training setup}
    Our training setup largely follows that of VIPT~\cite{anonymous2023cross} by updating the KLD regularization for prosody $\mathcal{L}_{ProsodyKLD}$ and noise $\mathcal{L}_{NoiseKLD}$ reference encoders with the phrase-level formulation and the length-based scaling coefficients $\beta$ described in Section~\ref{mehtod:length_based_regularization}. The final loss can be formulated as: 
    \begin{equation}
    \mathcal{L} = \mathcal{L}_{VITS} + \alpha_1 \mathcal{L}_{ProsodyKLD_{\beta_1}} + \alpha_2 \mathcal{L}_{NoiseKLD_{\beta_2}}
\end{equation}
where $\mathcal{L}_{VITS}$ represents the VITS loss terms with replaced adversarial components as in BigVGAN~\cite{lee2022bigvgan}. 
    We performed nine runs of hyperparameter search and set the KLD loss coefficient $\alpha$ and the length-based KLD loss scaling coefficient $\beta$ for both the prosody and the noise reference encoder as $\alpha_1=\alpha_2=0.04$ (other tested values: $0.02$ and $0.08$) and $\beta_1=\beta_2=0.08$ (other tested values: $0.02$ and $0.04$) respectively. We trained using mixed precision on 8 NVIDIA V100 GPUs, with a batch size of 30 per GPU, and used AdamW optimizer~\cite{loshchilovdecoupled}. We trained the model for 600 epochs. %
    The generative part of our proposed model and discriminators have 100 million and 47 million parameters respectively.

\subsection{Data}\label{experiments:data}
We used an internal multimedia dataset from which we extract multi-speaker multi-lingual dialogues resulting in 598 hours of speech from 134 female and 162 male speakers in 5 different locales; namely US English, Castilian Spanish, French, German and Italian. Speaker age groups range from children to elderly.

The speech data is resampled to 24 kHz and normalized in terms of loudness. Silences longer than 2 seconds are trimmed. We split the dialogues into separate phrases based on silences of at least 50 milliseconds.

\subsection{Evaluated systems}

We evaluated the proposed method against human Spanish dubs and two baseline models. We denote our method as Variational Inference for Prosody Transfer with Noise Modelling and Phrase-level Variational Auto-Encoder (VIPT-NM-PVAE). VIPT-NM-GVAE is a strong baseline for cross-lingual performance transfer with global-level reference encoder (corresponding to VIPT-NM-Transfer model from \cite{anonymous2023cross}). Additionally, we introduce a second baseline named VIPT-NM-GVAE-PP, which uses the same model architecture as VIPT-NM-GVAE during training, but at inference time it computes prosody embeddings per phrase (PP). Namely, during inference, the VIPT-NM-GVAE-PP model passes parts of the source audio corresponding to each of the $K$ phrases separately through the global-level reference encoder to extract the $K$ phrase-level embeddings. We include this baseline to evaluate the importance of training phrase-level embeddings. %

\begin{figure}[h]
  \centering
  \centerline{\includegraphics[width=\linewidth]{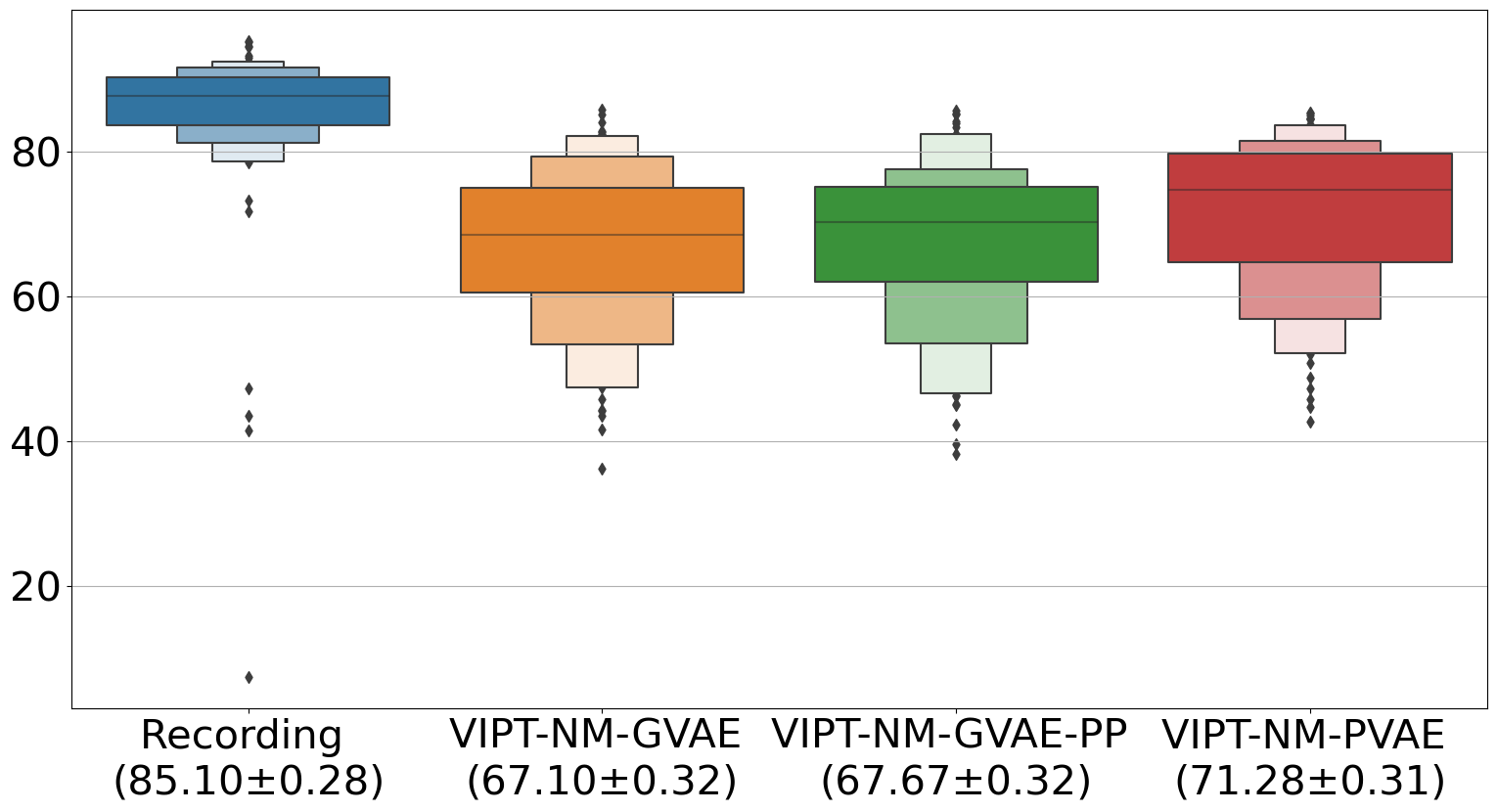}}
  \vspace{-2mm}
  \caption{Subjective listeners ratings from the machine dubbing MUSHRA test. Values under labels represent mean scores and their respective standard errors.}
  \label{fig:mushra}
\end{figure}
\vspace{-4mm}
\begin{table}[h]
\caption{\textit{Subjective evaluation MUSHRA mean scores reported separately for single-phrase and multi-phrase utterances.}}
\centering
\begin{tabular}{lll}
\toprule
\multirow{2}{*}{\textbf{System}}  &  \multicolumn{2}{c}{\textbf{MUSHRA $\uparrow$}} \\
& \multicolumn{1}{c}{single-phrase} & \multicolumn{1}{c}{multi-phrase} \\ \hline
\textbf{VIPT-NM-GVAE}            &  \multicolumn{1}{l}{$69.58 \pm 0.40$} & $63.21 \pm 0.52$    \\ %
\textbf{VIPT-NM-GVAE-PP}         &  \multicolumn{1}{l}{$69.81 \pm 0.39$} & $64.32 \pm 0.53$    \\ %
\textbf{VIPT-NM-PVAE}            &  \multicolumn{1}{l}{$\mathbf{74.31 \pm 0.36}$} & $\mathbf{66.56 \pm 0.53}$    \\ %
\textbf{Recording}                &  \multicolumn{1}{l}{$83.75 \pm 0.38$} & $87.22 \pm 0.40$    \\
\bottomrule 
\end{tabular}
\label{table:mushra}
\end{table}

\subsection{Subjective Evaluation}
For the evaluation of cross-lingual prosody transfer, we performed a MUSHRA test on a held-out subset of 100 parallel utterances between US English and Castilian Spanish.
To provide testers context for the assessment of prosody match, all audio samples were overlaid on the corresponding videos. 25 bi-lingual test subjects  native in Castilian Spanish and fluent in English were presented with the video samples in a random order side-by-side. The test subjects were tasked to \textit{``Rate the vocal performance in the Spanish video dubbing samples with respect to the English reference video''}. Each test case was scored by all 25 testers independently.

Evaluation results are summarized in Figure~\ref{fig:mushra} and show that VIPT-NM-PVAE achieved a statistically significant 6.2\% MUSHRA score increase over VIPT-NM-GVAE baseline system, which closes the gap to human dubbing by 23.2\%. 
Inspection of evaluators ratings %
suggests that the improvement in VIPT-NM-PVAE results from increased expressiveness of generated speech and more accurate prosody transfer. There is significant difference in MUSHRA scores between VIPT-NM-PVAE and both baseline models for multi-phrase utterances, and for single-phrase utterances (Table \ref{table:mushra}). We hypothesize that training with the proposed phrase-level reference encoder may lead to increased sensitivity to the prosody embedding.

\begin{table}[h]
\caption{\textit{Objective metrics: word error rate (WER) and conditional Fréchet DeepSpeech Distance (cFDSD) \cite{binkowski2019high}.}}
\centering
\begin{tabular}{llll}
\toprule
\multirow{2}{*}{\textbf{System}} & \multirow{2}{*}{\textbf{cFDSD $\downarrow$}} & \multicolumn{2}{c}{\textbf{WER $\downarrow$}} \\
& & \multicolumn{1}{c}{all} & \multicolumn{1}{c}{shortest 25\%} \\ \hline
\textbf{VIPT-NM-GVAE}            & $0.297$ & \multicolumn{1}{l}{$0.098$} & $0.169$     \\ %
\textbf{VIPT-NM-GVAE-PP}         & $0.288$ & \multicolumn{1}{l}{$0.094$} & $0.155$    \\ %
\textbf{VIPT-NM-PVAE}            & $\mathbf{0.224}$ & \multicolumn{1}{l}{$0.101$} & $0.161$     \\ %
\hspace{2mm}w/o length-based reg.           & $0.241$ & \multicolumn{1}{l}{$0.106$} & $0.229$    \\ %
\bottomrule 
\end{tabular}
\label{table:objective_metrics}
\vspace{-4mm}
\end{table}

\subsection{Objective Metrics}
To quantify stability of tested systems and intelligibility of synthesised speech we conducted Word Error Rate (WER) analysis. The results are reported in Table \ref{table:objective_metrics} for a held-out test set of 1200 parallel utterances. First, all generated audio files were transcribed with a Whisper Large \cite{radford2022robust} ASR model. Then, WER scores were computed between sentence texts and corresponding transcriptions. We have observed no significant stability issues with the VIPT-NM-PVAE model, which backs up our conclusion that phrase-level modelling allows for more expressive and accurate cross-lingual prosody transfer without compromising intelligibility.

For all tested systems we also computed the conditional Fréchet DeepSpeech Distance (cFDSD) \cite{binkowski2019high}, an objective metric measuring the quality of synthesised audio samples based on their distance to a reference set. We closely follow \cite{binkowski2019high} in our implementation of the cFDSD metric, only differing in using XLSR-53 Large \cite{conneau2020unsupervised} as a backbone network, which was trained on multi-lingual speech data. All tested systems are compared to human Spanish dubs. We observe that VIPT-NM-PVAE has a significantly lower distance to the human dubs (Table \ref{table:objective_metrics}) compared to all other models. This result is inline with the MUSHRA subjective evaluation scores.

Finally, as an ablation study, we trained a VIPT-NM-PVAE model without the length-based regularization described in Section \ref{mehtod:length_based_regularization}. This resulted in a significant WER increase for short utterances (Table \ref{table:objective_metrics}), while at the same time cFDSD distance to recordings also increased. We conclude that applying regularization dependent on phrase lengths is crucial to find a good balance between expressivity and stability of our system.

\section{Conclusions}
We have presented a novel solution that enables phrase-level cross-lingual, cross-speaker prosody transfer for expressive machine dubbing. The proposed method can learn to model prosody information at phrase-level, and transfer the phrase prosody embeddings from a source to a target language for translated text. In subjective evaluations, our system outperforms a strong baseline that transfers prosody at global-level. In future work, we plan to extend our evaluation to include wider range of languages and further close the gap between synthesised speech and expressive human dialogues by exploring duration modelling, hierarchical prosody modelling and the usage of parallel data.

\bibliographystyle{IEEEtran}
\bibliography{mybib}

\begin{thebibliography}{10}
\providecommand{\url}[1]{#1}
\csname url@samestyle\endcsname
\providecommand{\newblock}{\relax}
\providecommand{\bibinfo}[2]{#2}
\providecommand{\BIBentrySTDinterwordspacing}{\spaceskip=0pt\relax}
\providecommand{\BIBentryALTinterwordstretchfactor}{4}
\providecommand{\BIBentryALTinterwordspacing}{\spaceskip=\fontdimen2\font plus
\BIBentryALTinterwordstretchfactor\fontdimen3\font minus
  \fontdimen4\font\relax}
\providecommand{\BIBforeignlanguage}[2]{{%
\expandafter\ifx\csname l@#1\endcsname\relax
\typeout{** WARNING: IEEEtran.bst: No hyphenation pattern has been}%
\typeout{** loaded for the language `#1'. Using the pattern for}%
\typeout{** the default language instead.}%
\else
\language=\csname l@#1\endcsname
\fi
#2}}
\providecommand{\BIBdecl}{\relax}
\BIBdecl

\bibitem{googlee2e}
R.~Skerry-Ryan, E.~Battenberg, Y.~Xiao, Y.~Wang, D.~Stanton, J.~Shor, and
  R.~A.~S. Weiss, Ron J. Rob~Clark, ``{Towards End-to-End Prosody Transfer for
  Expressive Speech Synthesis with Tacotron},'' in \emph{International
  Conference on Machine Learning}.\hskip 1em plus 0.5em minus 0.4em\relax PMLR,
  2018.

\bibitem{wang2018style}
Y.~Wang, D.~Stanton, Y.~Zhang, R.-S. Ryan, E.~Battenberg, J.~Shor, Y.~Xiao,
  Y.~Jia, F.~Ren, and R.~A. Saurous, ``Style tokens: Unsupervised style
  modeling, control and transfer in end-to-end speech synthesis,'' in
  \emph{International Conference on Machine Learning}.\hskip 1em plus 0.5em
  minus 0.4em\relax PMLR, 2018, pp. 5180--5189.

\bibitem{luong2017adapting}
H.-T. Luong, S.~Takaki, G.~E. Henter, and J.~Yamagishi, ``Adapting and
  controlling dnn-based speech synthesis using input codes,'' in \emph{IEEE
  International Conference on Acoustics, Speech and Signal Processing
  (ICASSP)}.\hskip 1em plus 0.5em minus 0.4em\relax IEEE, 2017, pp. 4905--4909.

\bibitem{hsuhierarchical}
W.-N. Hsu, Y.~Zhang, R.~J. Weiss, H.~Zen, Y.~Wu, Y.~Wang, Y.~Cao, Y.~Jia,
  Z.~Chen, J.~Shen \emph{et~al.}, ``Hierarchical generative modeling for
  controllable speech synthesis,'' in \emph{International Conference on
  Learning Representations}, 2019.

\bibitem{8683623}
Y.-J. Zhang, S.~Pan, L.~He, and Z.-H. Ling, ``Learning latent representations
  for style control and transfer in end-to-end speech synthesis,'' in
  \emph{ICASSP 2019 - 2019 IEEE International Conference on Acoustics, Speech
  and Signal Processing (ICASSP)}, 2019, pp. 6945--6949.

\bibitem{guo2022emodiff}
Y.~Guo, C.~Du, X.~Chen, and K.~Yu, ``Emodiff: Intensity controllable emotional
  text-to-speech with soft-label guidance,'' \emph{IEEE International
  Conference on Acoustics, Speech and Signal Processing (ICASSP)}, 2023.

\bibitem{rattcliffe22_interspeech}
D.~Rattcliffe, Y.~Wang, A.~Mansbridge, P.~Karanasou, A.~Moinet, and M.~Cotescu,
  ``{Cross-lingual Style Transfer with Conditional Prior VAE and Style Loss},''
  in \emph{Proc. Interspeech 2022}, 2022, pp. 4586--4590.

\bibitem{Federico2020}
M.~Federico, R.~Enyedi, R.~Barra-Chicote, R.~Giri, U.~Isik, A.~Krishnaswamy,
  and H.~Sawaf, ``From speech-to-speech translation to automatic dubbing,'' in
  \emph{IWSLT 2020}, 2020.

\bibitem{matouvsek2012improving}
J.~Matou{\v{s}}ek and J.~V{\'\i}t, ``Improving automatic dubbing with subtitle
  timing optimisation using video cut detection,'' in \emph{2012 IEEE
  International Conference on Acoustics, Speech and Signal Processing
  (ICASSP)}.\hskip 1em plus 0.5em minus 0.4em\relax IEEE, 2012, pp. 2385--2388.

\bibitem{effendi2022duration}
J.~Effendi, Y.~Virkar, R.~Barra-Chicote, and M.~Federico, ``{Duration modeling
  of neural TTS for automatic dubbing},'' in \emph{ICASSP 2022-2022 IEEE
  International Conference on Acoustics, Speech and Signal Processing
  (ICASSP)}.\hskip 1em plus 0.5em minus 0.4em\relax IEEE, 2022, pp. 8037--8041.

\bibitem{anonymous2023cross}
J.~Swiatkowski, D.~Wang, M.~Babianski, G.~Coccia, P.~Lumban~Tobing,
  R.~Vipperla, V.~Klimkov, and V.~Pollet, ``Cross-lingual prosody transfer for
  expressive machine dubbing,'' in \emph{Interspeech}, 2023.

\bibitem{kim2021conditional}
J.~Kim, J.~Kong, and J.~Son, ``Conditional variational autoencoder with
  adversarial learning for end-to-end text-to-speech,'' in \emph{International
  Conference on Machine Learning}.\hskip 1em plus 0.5em minus 0.4em\relax PMLR,
  2021, pp. 5530--5540.

\bibitem{torresquintero2021adept}
A.~Torresquintero, T.~H. Teh, C.~G. Wallis, M.~Staib, D.~S.~R. Mohan, V.~Hu,
  L.~Foglianti, J.~Gao, and S.~King, ``{ADEPT: A Dataset for Evaluating Prosody
  Transfer},'' in \emph{Proc. Interspeech}, 2021, pp. 3880--3884.

\bibitem{renfastspeech}
Y.~Ren, C.~Hu, X.~Tan, T.~Qin, S.~Zhao, Z.~Zhao, and T.-Y. Liu, ``Fastspeech 2:
  Fast and high-quality end-to-end text to speech,'' in \emph{International
  Conference on Learning Representations}, 2020.

\bibitem{lee2019robust}
Y.~Lee and T.~Kim, ``Robust and fine-grained prosody control of end-to-end
  speech synthesis,'' in \emph{ICASSP 2019-2019 IEEE International Conference
  on Acoustics, Speech and Signal Processing (ICASSP)}.\hskip 1em plus 0.5em
  minus 0.4em\relax IEEE, 2019, pp. 5911--5915.

\bibitem{10022793}
M.~Babiański, K.~Pokora, R.~Shah, R.~Sienkiewicz, D.~Korzekwa, and V.~Klimkov,
  ``On granularity of prosodic representations in expressive text-to-speech,''
  in \emph{2022 IEEE Spoken Language Technology Workshop (SLT)}, 2023, pp.
  892--899.

\bibitem{9053520}
G.~Sun, Y.~Zhang, R.~J. Weiss, Y.~Cao, H.~Zen, and Y.~Wu, ``Fully-hierarchical
  fine-grained prosody modeling for interpretable speech synthesis,'' in
  \emph{ICASSP 2020 - 2020 IEEE International Conference on Acoustics, Speech
  and Signal Processing (ICASSP)}, 2020, pp. 6264--6268.

\bibitem{9414966}
Y.~Virkar, M.~Federico, R.~Enyedi, and R.~Barra-Chicote, ``Improvements to
  prosodic alignment for automatic dubbing,'' in \emph{ICASSP 2021 - 2021 IEEE
  International Conference on Acoustics, Speech and Signal Processing
  (ICASSP)}, 2021, pp. 7543--7574.

\bibitem{9747158}
J.~Effendi, Y.~Virkar, R.~Barra-Chicote, and M.~Federico, ``Duration modeling
  of neural tts for automatic dubbing,'' in \emph{ICASSP 2022 - 2022 IEEE
  International Conference on Acoustics, Speech and Signal Processing
  (ICASSP)}, 2022, pp. 8037--8041.

\bibitem{brannon2022dubbing}
W.~Brannon, Y.~Virkar, and B.~Thompson, ``Dubbing in practice: A large scale
  study of human localization with insights for automatic dubbing,''
  \emph{Transactions of the Association for Computational Linguistics}, 2021.

\bibitem{recommendation2015bs}
I.~Recommendation, ``Bs. 1534-3: Method for the subjective assessment of
  intermediate quality levels of coding systems,'' \emph{Geneva: International
  Telecommunications Union}, 2015.

\bibitem{binkowski2019high}
M.~Bi{\'n}kowski, J.~Donahue, S.~Dieleman, A.~Clark, E.~Elsen, N.~Casagrande,
  L.~C. Cobo, and K.~Simonyan, ``High fidelity speech synthesis with
  adversarial networks,'' \emph{International Conference on Learning
  Representations}, 2020.

\bibitem{Isik2020}
U.~Isik, R.~Giri, N.~Phansalkar, J.-M. Valin, K.~Helwani, and A.~Krishnaswamy,
  ``{PoCoNet: Better speech enhancement with frequency-positional embeddings,
  semi-supervised conversational data, and biased loss},'' in
  \emph{Interspeech}, 2020.

\bibitem{zhang2022visinger}
Y.~Zhang, J.~Cong, H.~Xue, L.~Xie, P.~Zhu, and M.~Bi, ``{VISinger: Variational
  inference with adversarial learning for end-to-end singing voice
  synthesis},'' in \emph{IEEE International Conference on Acoustics, Speech and
  Signal Processing (ICASSP)}.\hskip 1em plus 0.5em minus 0.4em\relax IEEE,
  2022, pp. 7237--7241.

\bibitem{cho2022sane}
H.~Cho, W.~Jung, J.~Lee, and S.~H. Woo, ``{SANE-TTS: Stable And Natural
  End-to-End Multilingual Text-to-Speech},'' in \emph{Proc. Interspeech}, 2022,
  pp. 1--5.

\bibitem{kong2020hifi}
J.~Kong, J.~Kim, and J.~Bae, ``{HiFi-GAN: Generative Adversarial Networks for
  Efficient and High Fidelity Speech Synthesis},'' \emph{Advances in Neural
  Information Processing Systems}, vol.~33, pp. 17\,022--17\,033, 2020.

\bibitem{lee2022bigvgan}
S.-g. Lee, W.~Ping, B.~Ginsburg, B.~Catanzaro, and S.~Yoon, ``{BigVGAN: A
  Universal Neural Vocoder with Large-Scale Training},'' \emph{International
  Conference on Learning Representations}, 2023.

\bibitem{Povey_ASRU2011}
D.~Povey, A.~Ghoshal, G.~Boulianne, L.~Burget, O.~Glembek, N.~Goel,
  M.~Hannemann, P.~Motlicek, Y.~Qian, P.~Schwarz, J.~Silovsky, G.~Stemmer, and
  K.~Vesely, ``The kaldi speech recognition toolkit,'' in \emph{IEEE 2011
  Workshop on Automatic Speech Recognition and Understanding}.\hskip 1em plus
  0.5em minus 0.4em\relax IEEE Signal Processing Society, Dec. 2011, iEEE
  Catalog No.: CFP11SRW-USB.

\bibitem{loshchilovdecoupled}
I.~Loshchilov and F.~Hutter, ``Decoupled weight decay regularization,'' in
  \emph{International Conference on Learning Representations}, 2019.

\bibitem{radford2022robust}
A.~Radford, J.~W. Kim, T.~Xu, G.~Brockman, C.~McLeavey, and I.~Sutskever,
  ``Robust speech recognition via large-scale weak supervision,'' \emph{arXiv
  preprint arXiv:2212.04356}, 2022.

\bibitem{conneau2020unsupervised}
A.~Conneau, A.~Baevski, R.~Collobert, A.~Mohamed, and M.~Auli, ``{Unsupervised
  cross-lingual representation learning for speech recognition},'' in
  \emph{Interspeech}, 2022.

\end{thebibliography}

\end{document}